\documentclass[fleqn,usenatbib]{mnras}

\usepackage{newtxtext,newtxmath}

\usepackage[T1]{fontenc}

\usepackage{graphicx}	
\usepackage{amsmath}	

\newcommand{\ergc}{\,erg\,cm$^{-2}$}	
\newcommand{\ergcc}{\,erg\,cm$^{-2}$\,cnt$^{-1}$}	
	
\title[Calibrating SPI-ACS/INTEGRAL]{Calibrating SPI-ACS/INTEGRAL for gamma-ray bursts and re-estimating energetics of GRB/GW 190425 in gamma-ray range}

\author[P. Yu. Minaev et al.]{
P. Yu. Minaev,$^{1}$\thanks{E-mail: minaevp@mail.ru}
A. S. Pozanenko,$^{1,2,3}$
\\
$^{1}$Space Research Institute of the Russian Academy of Sciences, Profsoyuznaya ul. 84/32, Moscow, 117997 Russia\\
$^{2}$National Research University “Higher School of Economics”, Myasnitskaya ul. 20, Moscow, 101000 Russia \\
$^{3}$Moscow Institute of Physics and Technology, Institutskiy per. 9, Dolgoprudny, 141701 Russia
}

\date{Accepted 2023 August 3, in original form 2023 June 23}

\pubyear{2023}

\begin{document}
\label{firstpage}
\pagerange{\pageref{firstpage}--\pageref{lastpage}}
\maketitle

\begin{abstract}
SPI-ACS/INTEGRAL is one of the most sensitive orbital gamma-ray detectors in energy range above 80 keV. Since 2002 it registered several thousands of gamma-ray bursts, including the bursts associated with LIGO-Virgo gravitational wave events GW 170817 and GW 190425. No dedicated in-flight calibrations were performed for  SPI-ACS/INTEGRAL, complicating estimation of spectral and energetic characteristics of an event. Using data of GBM/Fermi we perform cross-calibration of SPI-ACS/INTEGRAL, based on 1032 bright GRBs registered by both experiments. We find the conversion factor between instrumental counts from SPI-ACS and energy units from GBM to be dependent on hardness of GRB spectrum (defined as the characteristic energy value, $E_\text{p}$) and on location of a source in spacecraft based coordinate system. We determine the corresponding analytical model to calculate the conversion factor and estimate its accuracy empirically. Sensitivity of SPI-ACS/INTEGRAL to detect gamma-ray transients is also investigated. Using the calibration we re-estimate energetics of GRB/GW 190425, detected by SPI-ACS/INTEGRAL alone. We constrain possible range of the characteristic energy $E_\text{p}$ and isotropic equivalent of total energy, emitted in gamma-rays $E_\text{iso}$ for GRB 190425, using the $ E_\text{p,i} $ -- $ E_\text{iso} $  (Amati) correlation. The calibration model could be applied to any transients with energy spectrum, analogous to gamma-ray bursts. 
\end{abstract}

\begin{keywords}
gamma-ray burst: general -- methods: data analysis -- methods: statistical -- neutron star mergers
\end{keywords}



\section{Introduction}

Gamma-ray bursts (GRB) are among the most mysterious events in the Universe since their discovery in late 1960s by the U.S. Vela nuclear test detection satellites \citep{kleb73}. Bimodal duration distribution of bursts, discovered in a series of KONUS experiments forty years ago \citep{maz81} and confirmed later by a number of other experiments \citep[e.g.][]{deza92,kou93,min10b}, indicated presence of several types of their progenitors. 

Type I (short) bursts are associated with a merger of compact binaries, consisted of two neutron stars \citep[e.g.][]{blin84,pac86,mes92,ros03}. The connection was recently confirmed by LIGO-Virgo gravitational wave experiments for GRB 170817A, marking the beginning of a new era of multi-messenger astronomy \citep[e.g.][]{abb17,abb17b,gol17,sav17,poz18}. 

Type II (long) bursts are associated with a core collapse of massive stars \citep[e.g.][]{woo93,pac98,mes06}, confirmed by observations of bright (intrinsically) Ic supernovae, accompanying $\sim$ 10\% of type II GRBs with an identified optical component and a measured redshift \citep[e.g.][]{gal98,kul98,hjo03,can17,vol17,belk20}.

Detection of the second binary neutron star merger GW 190425 by LIGO-Virgo experiments \citep{abb20} revealed observational challenges in searching for the electromagnetic counterpart, because the localization area of the GW source was very large (LIGO Hanford Observatory (H1) was not taking data at the time, the event had low signal-to-noise ratio in Virgo). As a result, no success in a search of the burst afterglow and kilonova components in optical range was obtained \citep[e.g.][]{cou20}. 

The only electromagnetic counterpart, discovered for GW 190425, was gamma-ray burst GRB 190425, detected by SPI-ACS/INTEGRAL \citep{poz20}. The burst was not found in data of GBM/Fermi experiment, most probably due to  shielding of the source by the Earth. The non-detection of GRB 190425 by BAT/Swift could be connected with the shielding by the Earth or with the location of the source outside the BAT/Swift field of view. Worth to mention, GRB 170817A, associated with the gravitational wave event GW 170817, was also detected by SPI-ACS/INTEGRAL \citep{sav17,poz18,poz20}. 

Thus, the SPI-ACS experiment is the only experiment, detected both GRB 170817A and GRB 190425. It is actually one of the most sensitive gamma-ray detectors in energy range above 80 keV with effective area of about 0.3 m$^2$~\citep{kien03}. Since 2002 it registered several thousands of GRBs \citep[e.g.][]{rau05,min10a,sav12,min17,mozg21}. Due to high orbit of INTEGRAL observatory, SPI-ACS has several advantages over missions with low altitudes, such as Fermi and Swift (see Section~\ref{acs_descr}). The only significant disadvantage is the absence of dedicated in-flight calibrations, complicating estimation of spectral and energetic characteristics of an event. 

The calibration could be performed via Monte-Carlo simulations \citep[e.g.][]{sav17}, but the results of these simulations (responce matrices for different inclination angles and for different energy spectra of sources) are not publicly available, as well as the detailed mass-model of the spacecraft, necessary for the simulations. The simulations need also additional cross-calibration procedures to compensate for variance in properties of individual scintillation modules. 

Another method of SPI-ACS/INTEGRAL calibration, performed e.g. in \citet{vig09,poz20}, considers direct cross-calibration of the detector using spectral measurements of GRBs by other, well-calibrated experiments (e.g. GBM/Fermi, BAT/Swift, Konus/WIND). The method counts for instrumental effects, connected with a variance of individual scintillation modules, the mass-model of the spacecraft is not needed. The precision of the method depends crucially on size and quality of the GRB sample, used for the cross-calibration. In previous works the sample was relatively small (133 events in \citet{vig09} and 278 events in \citet{poz20}), so the detailed investigation of SPI-ACS sensitivity to GRBs with different energy spectra, registered from different inclination angles, was not performed.

In this paper, we perform the cross-calibration of SPI-ACS/INTEGRAL against GBM/Fermi, using the sample of 1032 bright GRBs registered by both detectors. We develop the detailed, publicly available, analytical model of conversion factor between instrumental counts from SPI-ACS and energy units from GBM, dependent on hardness of GRB spectrum (defined as the characteristic energy value, $E_\text{p}$) and coordinates of the source in spacecraft based coordinate system. We also estimate the sensitivity of SPI-ACS/INTEGRAL to detect different transients, depending on their duration, spectral hardness, location of the source in the spacecraft based coordinate system. Using the calibration model we re-estimate energetics of GRB 190425, revealing sufficiently more constrained estimates of isotropic equivalent of total energy, emitted in gamma-rays, $E_\text{iso}$ and the characteristic energy value, $E_\text{p}$.

The calibration is important groundwork for the O4 cycle of gravitational wave observations by LIGO-Virgo-KAGRA in 2023, when a registration of a GRB by SPI-ACS/INTEGRAL alone is also possible.

\begin{figure}
	\includegraphics[width=\columnwidth]{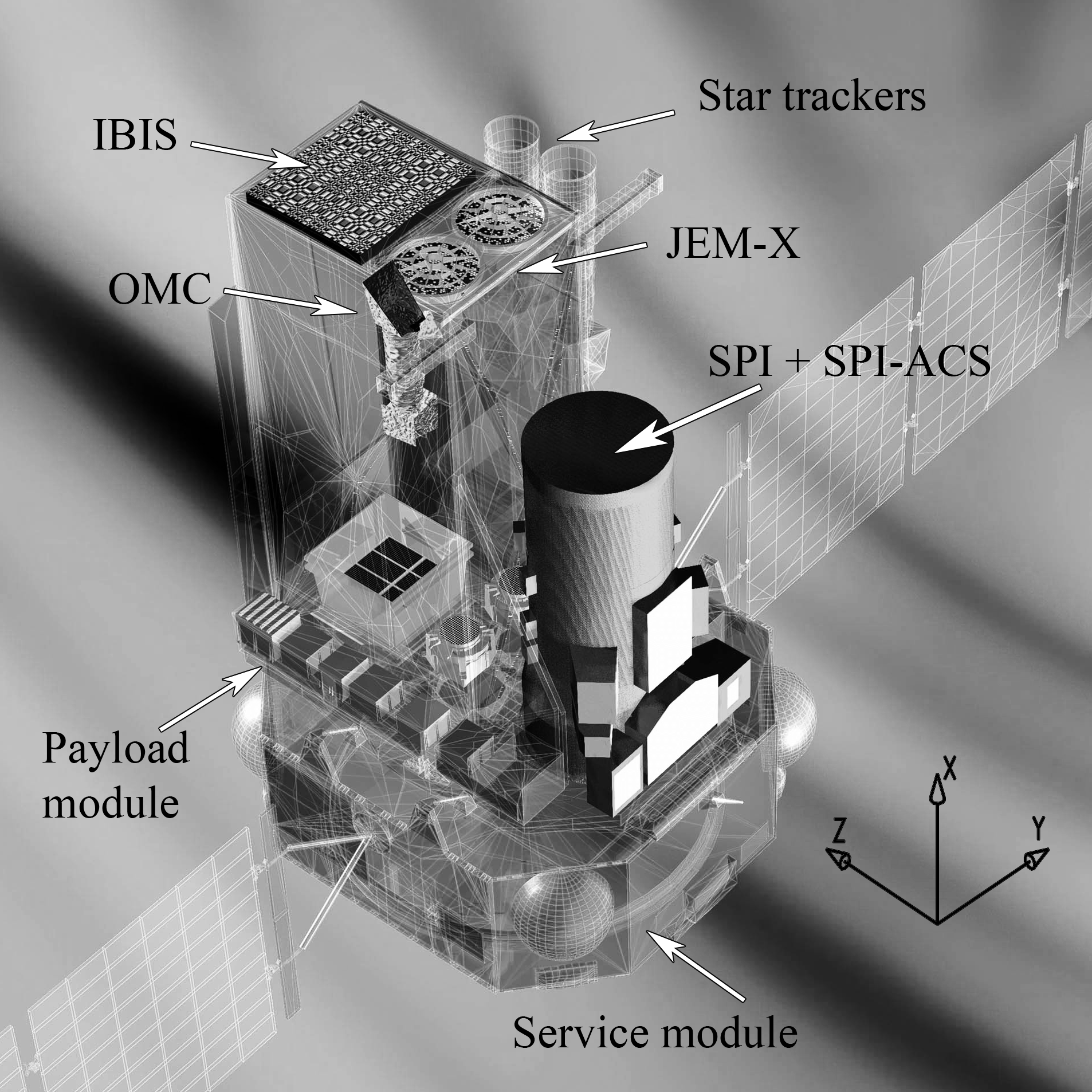}
    \caption{Location of instrumentation on the INTEGRAL spacecraft and spacecraft based coordinate system (image credit: ESA). }
    \label{fig:integral}
\end{figure}

\section{Calibration of SPI-ACS/INTEGRAL}

In this section we investigate the dependence of SPI-ACS/INTEGRAL sensitivity on spectral characteristics of gamma-ray bursts and on their location in spacecraft based coordinate system. The analytical model of conversion factor between SPI-ACS/INTEGRAL instrumental counts and energy units \ergc~in (10, 1000) keV energy range is developed.

\subsection{The SPI-ACS/INTEGRAL experiment}
\label{acs_descr}

The INTEGRAL observatory was launched into a highly elliptical orbit (the perigee and apogee of its initial orbit are 9000 and 153000 km, respectively) with a period of 72 h on October 17, 2002 \citep{jen03}. The observatory consists of the IBIS, SPI, JEM-X, OMC co-aligned apertured telescopes (Fig.~\ref{fig:integral}).

SPI-ACS is the anticoincidence shield of the SPI telescope, composed of 91 bismuth germanate (BGO) crystals with a total mass of 512 kg and a maximum effective area of about 0.3 m$^2$. It is used to reduce the background of the SPI detectors associated with the interaction of the equipment with cosmic rays \citep{kien03}. The thickness of the individual crystals ranges from 16 to 50 mm. They are positioned in two rings (the upper and lower collimator ring) whose axes are along the viewing direction of the spectrometer, between the coded mask and the detector plane of the SPI; in addition, there are side-shield and rear-shield assemblies \citep{rau05}. In addition to the BGO, a plastic scintillator (PSAC), whose purpose is the reduction of the 511 keV background produced by particle interactions in the passive mask, is located directly below the SPI coded mask \citep{rau05}.

\begin{figure*}
	\includegraphics[width=2\columnwidth]{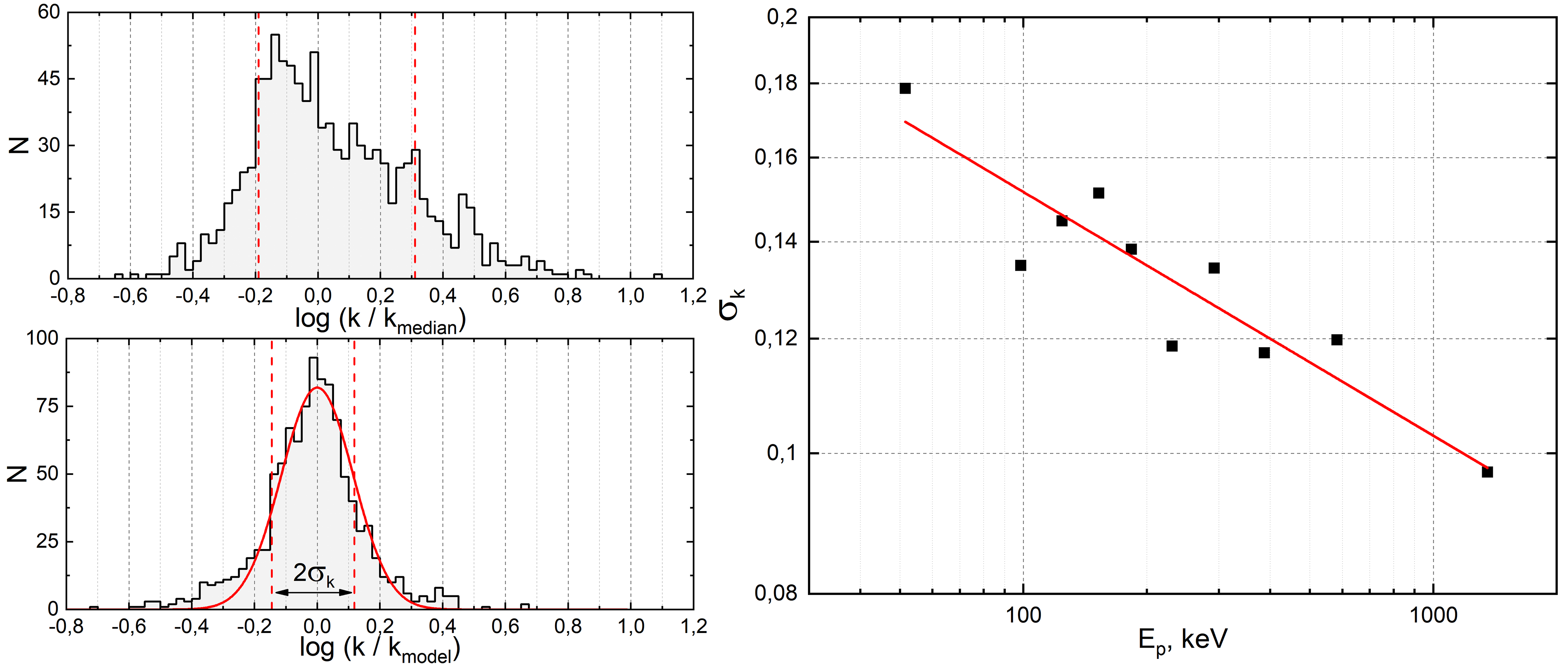}
    \caption{Left: distribution of deviation of a measured conversion factor $k$ value from the median value of $k_\text{median}$ = 2.37 $\times~ 10^{-10}$\ergcc (top),  distribution of deviation of a measured $k$ value from the analytical model $k_\text{model}$ (bottom) for the whole sample of 1032 GRBs. Dashed red vertical lines, binding 68\% percentiles of the distributions, represent the width of them and the nominal standard deviation of the conversion factor $\sigma_\text{k}$, defined as the half-width of corresponding distribution. Red curve in the bottom figure shows gaussian fit of the distribution.   
    Right: dependence of standard deviation $\sigma_\text{k}$ on the characteristic energy $E_\text{p}$ (spectral hardness) for 10 subsamples, red line shows power-law fit.}
    \label{fig:scatter}
\end{figure*}

Each of the 90 BGO crystals is viewed by two photomultipliers (PMTs), which are read out by 90 front-end electronic (FEE) boxes (The 91th BGO crystal is read out by a single PMT and FEE). The counts from all units are recorded in a single channel with time resolution of 50 ms (spectral information about events is not available). SPI-ACS provides a quasiomnidirectional field of view. A lower sensitivity threshold of the detector is $\sim$ 80 keV -- the physical properties of the individual BGO assemblies (detector + PMT + FEE) slightly differ and, therefore, have different lower energy thresholds. Upper limit to the energy range is not defined and exceeds 10 MeV \citep{kien03}. 

High orbit of the INTEGRAL observatory gives several advantages comparing with low orbit observatories with altitude of $\sim$ 600 km, such as Fermi, Swift and RHESSI. Firstly, the Earth occupies an insignificant area in the sky without limiting the field of view. Secondly, the background signal is stable at higher timescales: in SPI-ACS experiment it could be described by a linear model up to a few thousands of seconds \citep{mozg21}.

\subsection{Sample selection}

Our initial sample of GRBs, registered by SPI-ACS/INTEGRAL since 2002 and confirmed by other gamma-ray experiments, consists of almost four thousands events. The construction of the initial sample and its detailed analysis will be published in our other forthcoming paper. In this work we use only  fluence (time integrated energy flux) values of these GRBs, calculated in instrumental counts of the SPI-ACS detector. 

Next, we find coincidental events in the latest spectral GRB catalog of GBM/Fermi experiment \citep{pool21}. From the catalog we exclude: 1) weak GRBs with energy spectrum, best fitted by simple power-law model, 2) GRBs, which sources are occulted by the Earth during the burst (at least, partially).

After all matching and filtering procedures we obtain the final sample of 1032  bursts, registered by both SPI-ACS and GBM experiments, which  we use in our main analysis.

\subsection{Construction of the calibration model}

\subsubsection{Conversion factor, $k$}

In the GBM/Fermi catalog of gamma-ray bursts fluence is obtained in energy range of (10, 1000) keV \citep{pool21}. It does not match the energy range of SPI-ACS/INTEGRAL of (80, 10000) keV (Section~\ref{acs_descr}). The mismatch could give additional uncertainties in calibration, especially for events with soft energy spectrum ($E_\text{p}$ < 100 keV). Nevertheless, we use the energy range of (10, 1000) keV for the calibration because it is more commonly used in GRB science, than range of (80, 10000) keV. Our empirical method of estimation of calibration accuracy counts these mismatch effects  (Section~\ref{kvar}).

We define a conversion factor $k$ between instrumental counts and energy units via equation~\ref{eq:k}, where $F_\text{GBM}$ is a fluence of an event, obtained in GBM/Fermi experiment in energy range of (10, 1000) keV and expressed in units of \ergc, $F_\text{SPI-ACS}$ is a fluence of an event obtained in SPI-ACS/INTEGRAL experiment expressed in instrumental counts.

\begin{equation}
    k = \frac{F_\text{GBM}}{F_\text{SPI-ACS}} 
	\label{eq:k}
\end{equation}

We calculate the conversion factor for the whole sample of 1032 events and obtain its median value as $k_\text{median}$ = 2.37 $\times~ 10^{-10}$\ergcc. The scatter of the $k$ values is found to be large and covers more than 1.5 orders of magnitude. The distribution of the scatter is presented in top-left part of Fig.~\ref{fig:scatter}. The distribution is highly asymmetric with maximum placed at $k_\text{peak}$ $\simeq$ 0.8 $k_\text{median}$. We calculate 68\% percentiles for the distribution (red dashed vertical lines at Fig.~\ref{fig:scatter}), which bind the nominal 1$\sigma$ confidence region of $\frac{k}{k_\text{median}}$ as (0.65, 2.04).

Large width and asymmetry of the $k$ distribution are connected (at least partially) with dependence of SPI-ACS/INTEGRAL sensitivity on spectral characteristics of an event and on location of its source in spacecraft based coordinate system. The value of the conversion factor anti-correlates with the sensitivity (effective area) of SPI-ACS instrument: larger effective area results in higher amount of instrumental counts (fluence) for a GRB, and a smaller $k$ value as consequence (equation~\ref{eq:k}).

\begin{figure*}
	\includegraphics[width=2\columnwidth]{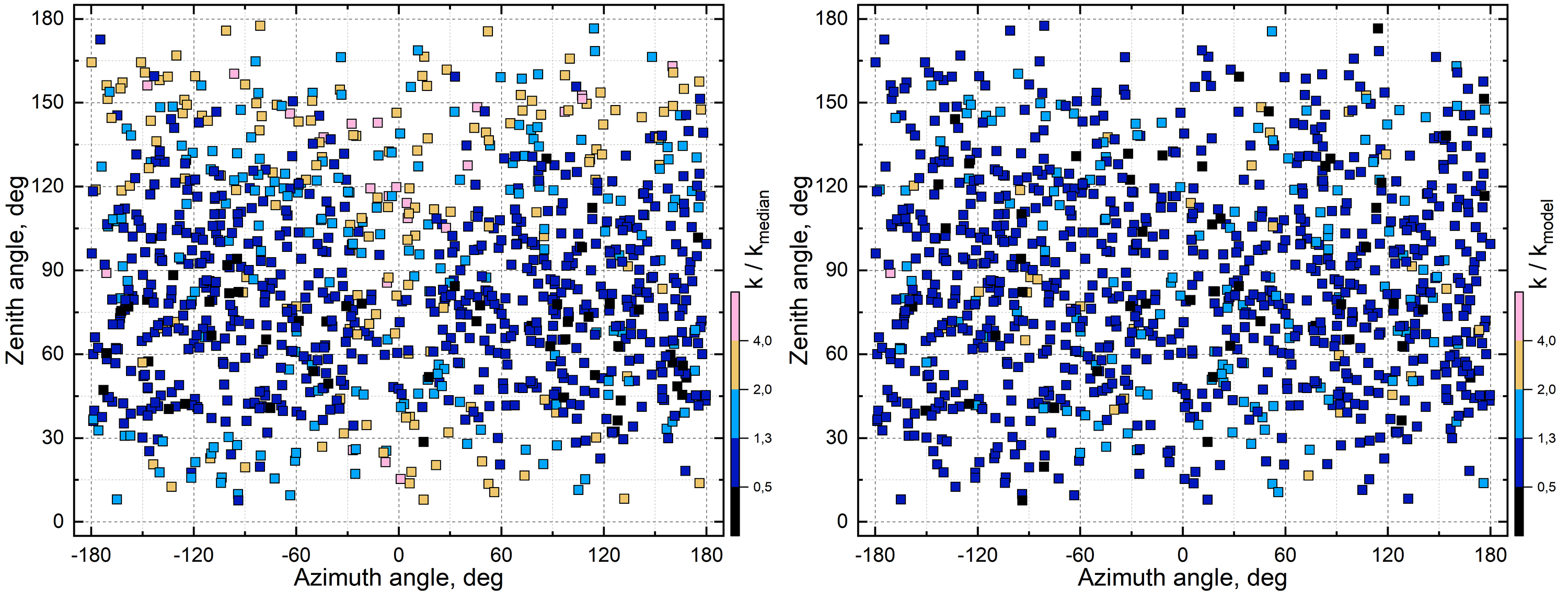}
    \caption{Dependence of conversion factor $k$ on azimuth and zenith angles ($a$, $z$) in spacecraft based coordinate system before corrections (left) and after all corrections (right). Each symbol corresponds to a single GRB of the whole sample of 1032 events. Its colour represents the ratio between the measured $k$ value and the median one $k_\text{median}$ = 2.37 $\times~ 10^{-10}$\ergcc (left), and the ratio between the measured value and the analytical model $k_\text{model}$, calculated via equation~\ref{eq:kmodel} (right). Corresponding colour tables are shown in bottom right part of each figure.  }
    \label{fig:2D}
\end{figure*}

The significant dependence of conversion factor $k$ on the location of a GRB source in spacecraft based coordinate system is seen in the left part of Fig.~\ref{fig:2D}. The decrease of the SPI-ACS effective area in certain directions results also in marginal decrease of a events number density in these directions, also visible in Fig.~\ref{fig:2D}.

Firstly, the effective area of SPI-ACS depends on the zenith angle $z$, the angle between the center of field of view of INTEGRAL instruments (aligned along X axis, Fig.~\ref{fig:integral}) and the direction to the GRB source ($z$ = 0\degr~for the source, located in the center of field of view of INTEGRAL instruments). The effective area decreases towards directions $z$ = 0\degr~and $z$ = 180\degr~due to several factors: 1) the decreasing of the corresponding geometrical area -- SPI-ACS could be roughly described as cylinder  ($z$ = 0\degr~and $z$ = 180\degr~correspond to the illumination of top and bottom surfaces of the cylinder, respectively); 2) plastic scintillator PSAC, placed at the top surface of SPI-ACS, is more sensitive to charged particles, but not the gamma rays (comparing with BGO crystals, mounted at the sides of the cylinder); 3) INTEGRAL Service module (Fig.~\ref{fig:integral}), placed at the bottom of the observatory ($z$ $\sim$ 180\degr), shields the source partially.

Secondly, the effective area of SPI-ACS depends on the azimuth angle $a$, the angle between the direction of Z axis (Fig.~\ref{fig:integral}) and the projection of the direction to the GRB source in YZ plane ($a$ = 0\degr~for the source located in XZ plane with IBIS instrument on the way between the source and SPI-ACS). The effective area of SPI-ACS decreases at $a$ = 0\degr~probably due to shielding by IBIS and other instruments, while the amount of the decrease depends also on zenith angle: the absorption of gamma-ray emission is more pronounced probably at directions towards the service module, detectors (ISGRI and PICsIT) and coded mask of IBIS telescope.

In following subsections we estimate corresponding corrections to the conversion factor $k$.

\begin{figure}
	\includegraphics[width=\columnwidth]{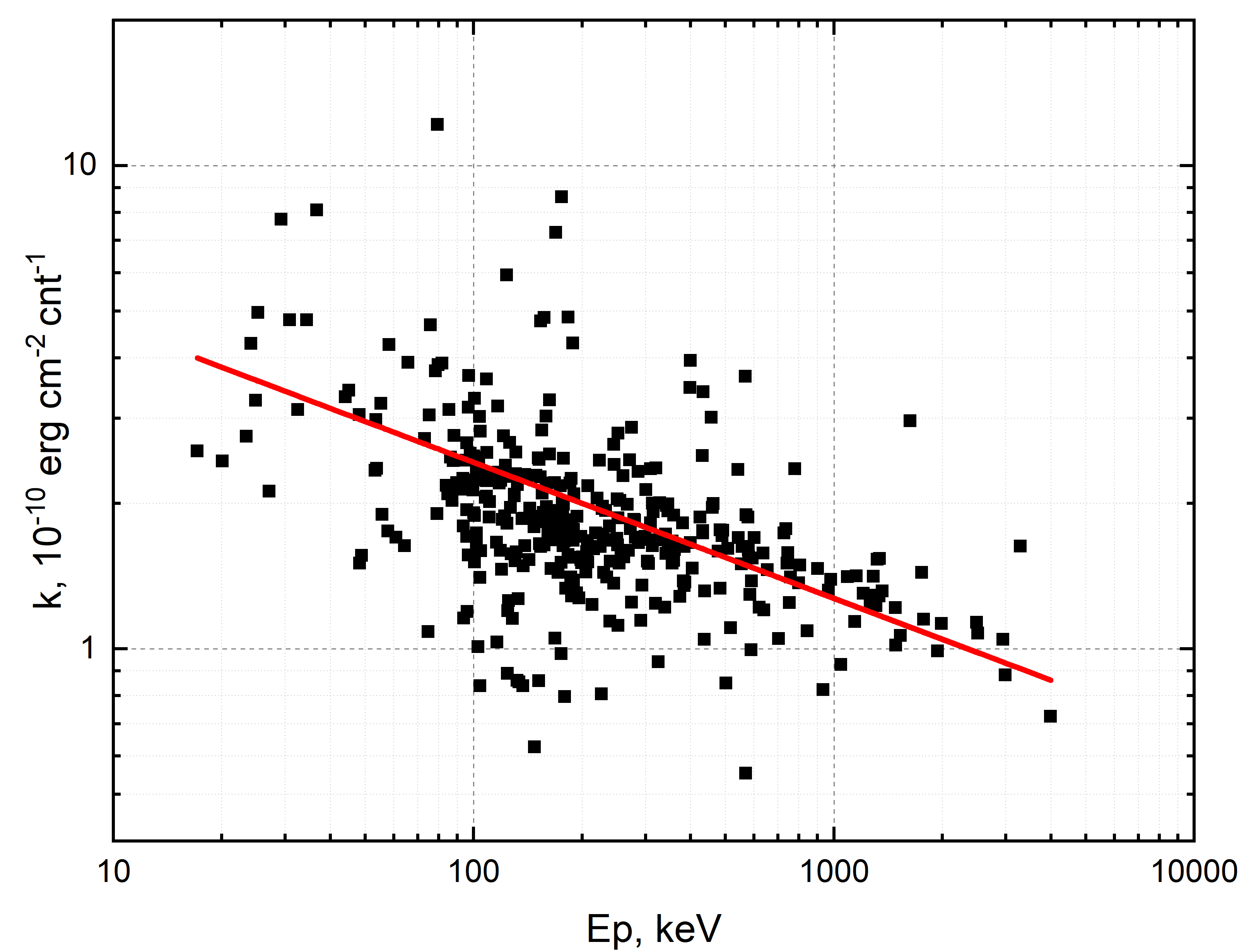}
    \caption{Dependence of conversion factor $k$ on characteristic energy $E_\text{p}$ for subsample of 368 GRBs with $z$ in range of (40\degr, 90\degr) and $|a|$ > 40\degr. Red line shows power-law fit (equation~\ref{eq:ke}).}
    \label{fig:ecor}
\end{figure}

\subsubsection{Spectral hardness ($E_\text{p}$) correction}

At first, we estimate correction, based on spectral characteristics of gamma-ray bursts. We consider the spectral correction to be not dependent on coordinates of the source in spacecraft based coordinate system, which is not correct in a general case (to investigate the dependence we need sufficiently larger sample of GRBs).

\begin{figure*}
	\includegraphics[width=2\columnwidth]{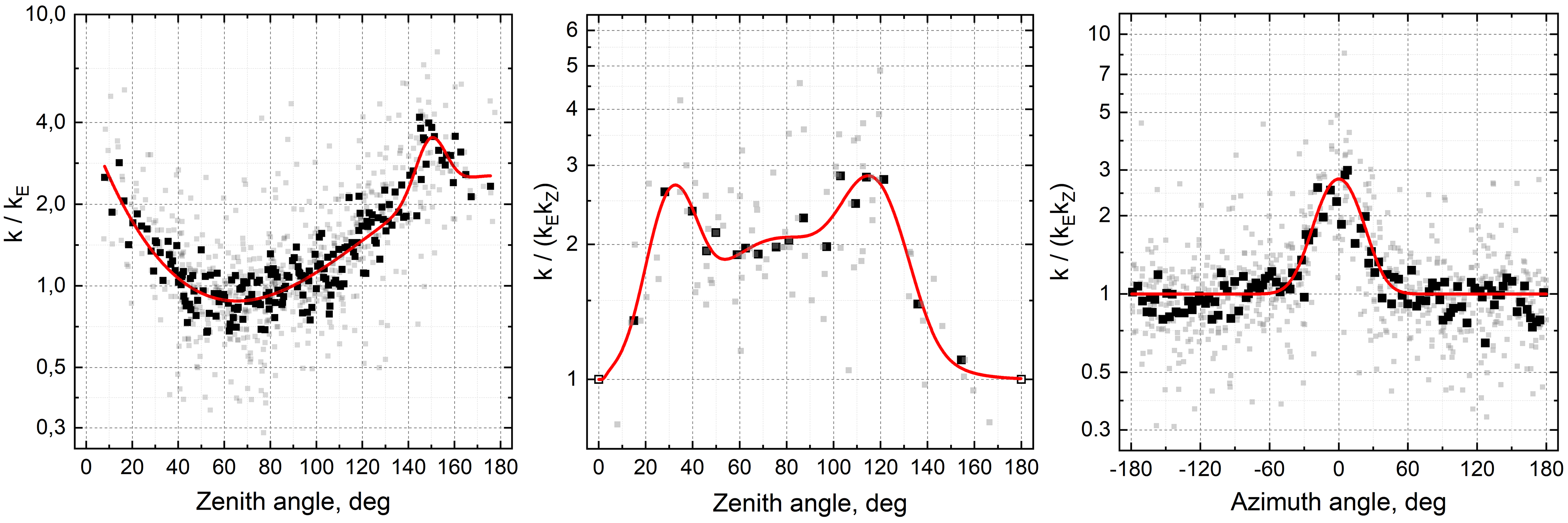}
    \caption{Left: dependence of conversion factor $k$ on zenith angle $z$ for subsample of 844 GRBs with azimuth anlge $|a|$ > 80\degr after spectral hardness correction $k_\text{E}$ (equation~\ref{eq:ke}), red curve represents the basic zenith correction fit (equation~\ref{eq:kez}). Center: dependence of conversion factor $k$ on zenith angle $z$ for subsample of 91 GRBs with azimuth angle $|a|$ < 25\degr after spectral hardness $k_\text{E}$ (equation~\ref{eq:ke}) and basic zenith angle $k_\text{Z}$ (equation~\ref{eq:kez}) corrections, red curve represents the additional zenith correction fit (equation~\ref{eq:kezsz}). Right: dependence of conversion factor $k$ on azimuth angle $a$ for subsample of 789 GRBs with zenith angle $z$ in range of (30\degr, 125\degr) after spectral hardness $k_\text{E}$ (equation~\ref{eq:ke}) and basic zenith angle $k_\text{Z}$ (equation~\ref{eq:kez}) corrections, red curve represents the additional azimuth correction fit (equation~\ref{eq:kezsa}). Small gray symbols show single GRB measurements, large black symbols are obtained using median filter (without overlap) with width of 5 events (left and center) and 7 events (right) to minimize impact of outliers. Additional unfilled symbols with $z$ = 0\degr~and $z$ = 180\degr~in the central figure represent required match of additional zenith correction with the basic zenith correction at poles of the coordinate system.}
    \label{fig:ccor}
\end{figure*}

As shown in the previous section, conversion factor $k$ depends on the coordinates of the GRB source. Therefore, for the spectral calibration we selected 368 events with $z$ in range of (40\degr, 90\degr) and $|a|$ > 40\degr, where the dependence is minimal (Figs.~\ref{fig:2D},\ref{fig:ccor}).

As we exclude from our sample weak bursts with low S/N ratio, whose energy spectrum is best fitted by simple power-law model, our final sample consists of events with energy spectra described by three possible models: power-law with exponential cutoff (COMP), power-law with break (BAND), smoothly broken power-law (SBPL) \citep{pool21}. For the selected subsample of 368 events we investigate possible correlations of the conversion factor with different parameters of spectral model (power-law indices and characteristic energy $E_\text{p}$). 

The only statistically significant correlation found is the correlation of conversion factor with the characteristic energy $E_\text{p}$ (Fig.~\ref{fig:ecor}). We find the Spearman rank-order correlation coefficient $\rho$ = -0.5 and the associated null-hypothesis (chance) probability $P_\rho$ = 5 $\times$ $10^{-25}$ for the correlation. The dependence is fitted by power-law model (equation~\ref{eq:ke}), parameters of the model are listed in Table~\ref{ta_mopar}.

\begin{equation}
    k_\text{E} = A_1E_\text{p}^{~A_2}
	\label{eq:ke}
\end{equation}

The characteristic energy $E_\text{p}$ corresponds to the position of the extremum (maximum) in energy spectrum $\nu F_\nu$, i.e. it indicates spectral hardness of the burst. Low value of $E_\text{p}$ (e.g. below 100 keV) means the fluence of the burst is concentrated at low energies, where the efficiency of SPI-ACS detector is low, resulting in higher value of conversion factor $k$. Another effect, possibly leading to the $k$ -- $E_\text{p}$ correlation, is that single high energy photon could be detected by several BGO crystals (e.g. via Compton scattering), raising the total count rate.

In the next section we use spectral corrected value of conversion factor $k/k_\text{E}$ to estimate basic zenith angle correction.

\subsubsection{Basic zenith angle correction}

We divide correction, based on a location of a burst in spacecraft coordinate system ($a$, $z$), into two components. The first, basic one, mainly connected with geometric effects, we suppose to be not dependent on azimuth angle $a$. The second, additional correction is probably connected with shielding of SPI-ACS by other instruments (by IBIS telescope, mostly), which is dependent on both azimuth and zenith angles, and concentrated around $a$ = 0\degr~(the direction towards the IBIS telescope). 

In this subsection we estimate the basic zenith angle correction. We select 844 events in range of $|a|$ > 80\degr, where shielding effects are minimal (see Fig.~\ref{fig:2D}). The dependence of spectral corrected conversion factor $k/k_\text{E}$ on zenith angle for the subsample is shown in left part of Fig.~\ref{fig:ccor}. The scatter of the dependence for individual events is quite high, so we applied median filter with step of 5 events to the subsample without overlap (a single event is used only for a single median filter bin). 

The dependence is fitted in logarithmic space by polynomial model with additional gaussian, describing feature at $z$ $\simeq$ 150\degr~(equation~\ref{eq:kez}, Table~\ref{ta_mopar}). Contribution of the feature could be dependent on azimuth angle, but we find no statistical evidence, probably due to insufficient number of events in the sample. The nature of the feature is unknown, it could be connected with shielding of SPI-ACS by Service module of the observatory (Fig.~\ref{fig:integral}).

\begin{equation}
   \lg(k_\text{Z}) = B_1+B_2z+B_3z^2+B_4z^3+B_5\exp\Big(-\frac{(z-B_6)^2}{2B_7^{~2}}\Big)
	\label{eq:kez}
\end{equation}

In the next subsection we obtain additional corrections, using values of conversion factor $k/(k_\text{E}k_\text{Z}$), corrected for spectral hardness and basic zenith angle effects.

\subsubsection{Additional zenith and azimuth angle correction}

We consider this correction (mostly connected with shielding of SPI-ACS by other telescopes and instrumentation of INTEGRAL) as additive one, consisting of two independent components:  $k_\text{SA}$ -- additional azimuth correction, $k_\text{SZ}$ -- additional zenith correction. The final analytical model, considering all discussed corrections, is calculated via equation~\ref{eq:kmodel}. 

\begin{equation}
    k_\text{model}(E_\text{p}, a, z) = k_\text{E}*k_\text{Z}*(1+k_\text{SA}*k_\text{SZ})
    \label{eq:kmodel}
\end{equation}

As seen in Fig.~\ref{fig:2D}, the width of the additional 'shielding' component does not depend significantly on zenith angle (except widening towards poles of coordinates system, $z$ = 0\degr~and $z$ = 180\degr). We select 789 events with $z$ in range of (30\degr, 125\degr), where deviation from basic zenith correction is significant and the widening towards poles is not significant. The corresponding dependence of conversion factor $k/(k_\text{E}k_\text{Z}$) on azimuth angle is shown in right part of Fig.~\ref{fig:ccor}. We apply median filter to the data with step of 7 events without overlap to reduce scatter. The dependence is symmetric and fitted by gaussian model with widening towards poles of the coordinate system, described by a factor of $\sin(z)$  (equation~\ref{eq:kezsa}, Table~\ref{ta_mopar}). 

\begin{equation}
   k_\text{SA} = \exp\Big(-\frac{[a\sin(z)]^2}{2C_1^{~2}}\Big)
	\label{eq:kezsa}
\end{equation}

We consider $k_\text{SA}$ component as the dependence of $k_\text{SZ}$ component on azimuth angle, therefore the amplitude of the gaussian, describing $k_\text{SA}$ component is set to 1. Actual amplitude of the additional correction is estimated in $k_\text{SZ}$ model of additional zenith correction.

To estimate additional zenith correction we formed subsample of 91 events with $a$ in range of (-25\degr, 25\degr), where the most part of the additional component is placed (see Fig.~\ref{fig:2D}). Corresponding dependence of conversion factor $k/(k_\text{E}k_\text{Z}$) on zenith angle is shown in central part of Fig.~\ref{fig:ccor}. As in case of basic zenith correction, we applied median filter to the data with step of 5 events without overlap. We also added two additional points at poles of the coordinate system, where the additional correction should match the basic zenith angle and spectral hardness correction ($k_\text{model}$ $\equiv$ $k/(k_\text{E}k_\text{Z}$) at $z$ = 0\degr~and $z$ = 180\degr).

The dependence is complicated and consists of two sharp peaks at $z$ $\simeq$ 30\degr~and $z$ $\simeq$ 115\degr. As we need additive model of $k_\text{SZ}$, we used sum of three gaussians, describing all discovered features. The model naturally tends to zero value at high zenith angles, but at low zenith angles the tendency is insufficient, so we added exponential growth to the model to ensure the condition $k_\text{model}$ $\equiv$ $k/(k_\text{E}k_\text{Z}$) at $z$ = 0\degr~(equation~\ref{eq:kezsz}, Table~\ref{ta_mopar}).    

\begin{equation}
   k_\text{SZ} = D_0\exp\Big(-\frac{10}{z^2}\Big)\sum_{i=1}^{3}\Big[D_1^i\exp\Big(-\frac{(z-D_2^i)^2}{{2D_3^i}^2}\Big)\Big]
	\label{eq:kezsz}
\end{equation}

Additional coefficient $D_0$ is calculated via equation~\ref{eq:d0}. The coefficient compensates the effect of underestimation of the amplitudes $D_1^i$ of the additional zenith correction (equation~\ref{eq:kezsz}), caused by investigation of subsample with relatively wide range of azimuth angles ($a_i$, $a_f$) = (-25\degr, 25\degr), while the amplitude of the correction gradually decreases with increase of $|a|$ (equation~\ref{eq:kezsa}).

\begin{equation}
  D_0 = \frac{a_f - a_i}{\int_{a_i}^{a_f} \exp\Big(-\frac{a^2}{2C_1^{~2}}\Big)\,da} \simeq 1.32
	\label{eq:d0}
\end{equation}

Thus, we obtain all necessary corrections to conversion factor $k$, which could be calculated as $k_\text{model}$ via equation~\ref{eq:kmodel}. The dependence of corrected conversion factor $k/k_\text{model}$ on azimuth and zenith angles for the whole sample of investigated GRBs is shown in right part of Fig.~\ref{fig:2D}.

\begin{figure}
	\includegraphics[width=\columnwidth]{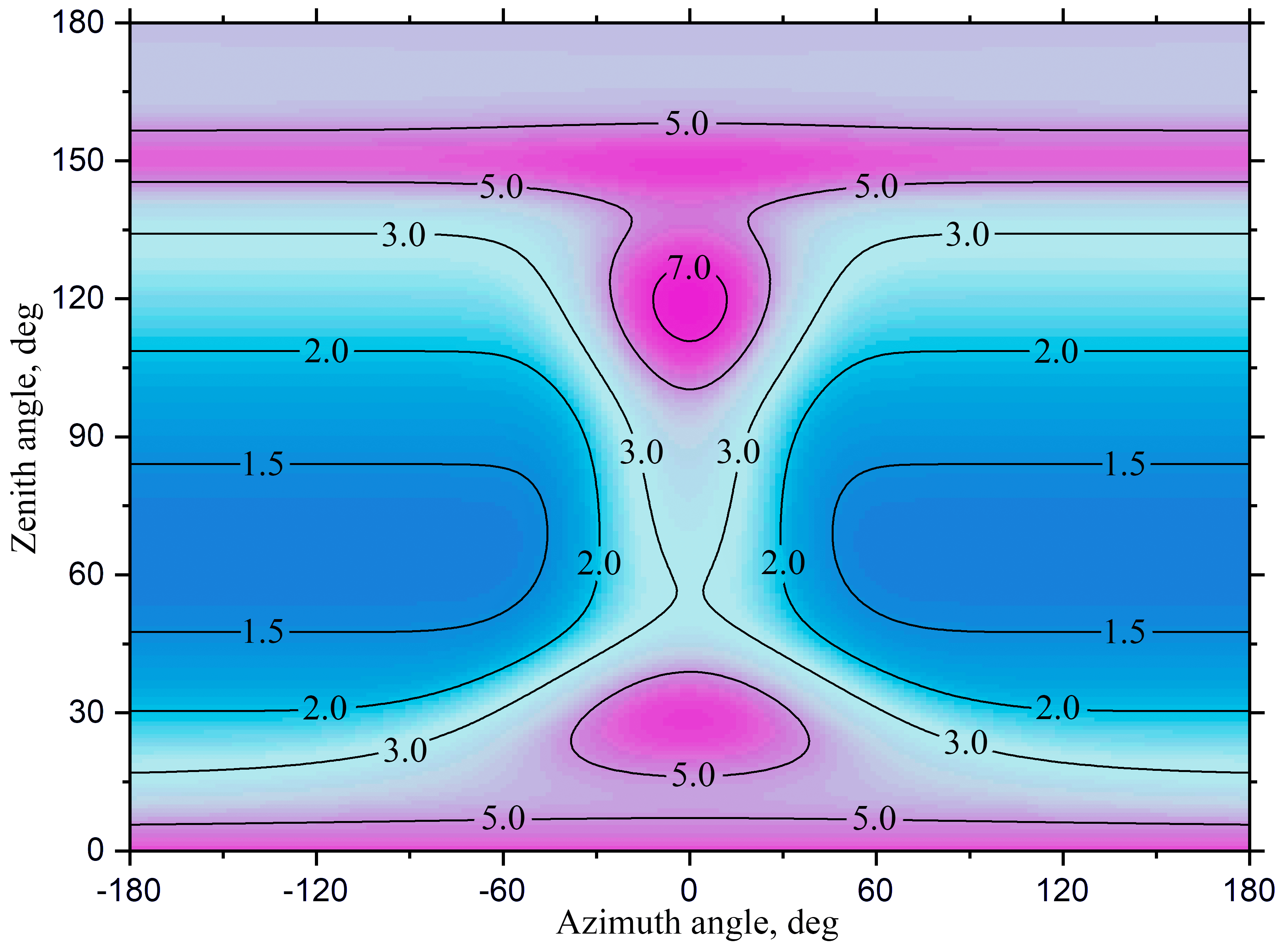}
    \caption{Dependence of conversion factor $k_\text{model}$ (equation~\ref{eq:kmodel}) on azimuth and zenith angle for a GRB with characteristic energy of $E_\text{p}$ = 500 keV. Labels on contour lines show the values of $k_\text{model}$ in units of $10^{-10}$\ergcc. }
    \label{fig:model}
\end{figure}

Fig.~\ref{fig:model} shows the model for a GRB with characteristic energy of $E_\text{p}$ = 500 keV. The value of conversion factor $k_\text{model}$ varies in wide range from $k_\text{model}$ $\simeq$ 1.4 $\times$ $10^{-10}$\ergcc~upto $k_\text{model}$ $\simeq$ 7.8 $\times$ $10^{-10}$\ergcc, confirming the necessity of azimuth and zenith corrections.

\begin{table*}

 \caption{Parameters of the analytical model $k_\text{model}$ of conversion factor.}  \label{ta_mopar}

 \begin{tabular}{lccccccc}
  \hline

Index$^a$ &	$k_\text{E}$  &	$k_\text{Z}$	&	$k_\text{SA}$ & $k_\text{SZ}^1$ & $k_\text{SZ}^2$ & $k_\text{SZ}^3$  & $\sigma_\text{k}$\\	\hline
1  & $(8.9 \pm 1.3)\times10^{-10}$ & 0.61 $\pm$ 0.04 & 18.14 $\pm$ 0.03 &  1.38 $\pm$ 0.29 & 1.07 $\pm$ 0.11 & 1.35 $\pm$ 0.51 & 0.33 $\pm$ 0.05 \\	
2  & $ -0.28 \pm 0.03$ & (-2.32 $\pm$ 0.18)$\times10^{-2}$ & - &  31.5 $\pm$ 1.6  & 78.1 $\pm$ 11.2 & 117.0 $\pm$ 2.0 & -0.17 $\pm$ 0.03\\	
3  & - & (2.43 $\pm$ 0.22)$\times10^{-4}$ & - & 9.3 $\pm$ 2.1 & 30.7 $\pm$ 13.3 & 12.1 $\pm$ 3.3 & - \\	
4  & - & (-6.67 $\pm$ 0.81)$\times10^{-7}$ & - &  - & - & - & -\\	
5  & - & 0.208 $\pm$ 0.034 & - &  - & - & -  & -\\	
6  & - & 149.6 $\pm$ 1.2 & - &  - & - & -  & -\\	
7  & - & 6.55 $\pm$ 1.42 & - &  - & - & -  & -\\	
\hline

\multicolumn{7}{l}{$^a$ - parameter index in the corresponding model. $k_\text{SZ}$ model is divided in to three components (see equation~(\ref{eq:kezsz}).} 

 \end{tabular}
\end{table*}

\begin{figure}
	\includegraphics[width=\columnwidth]{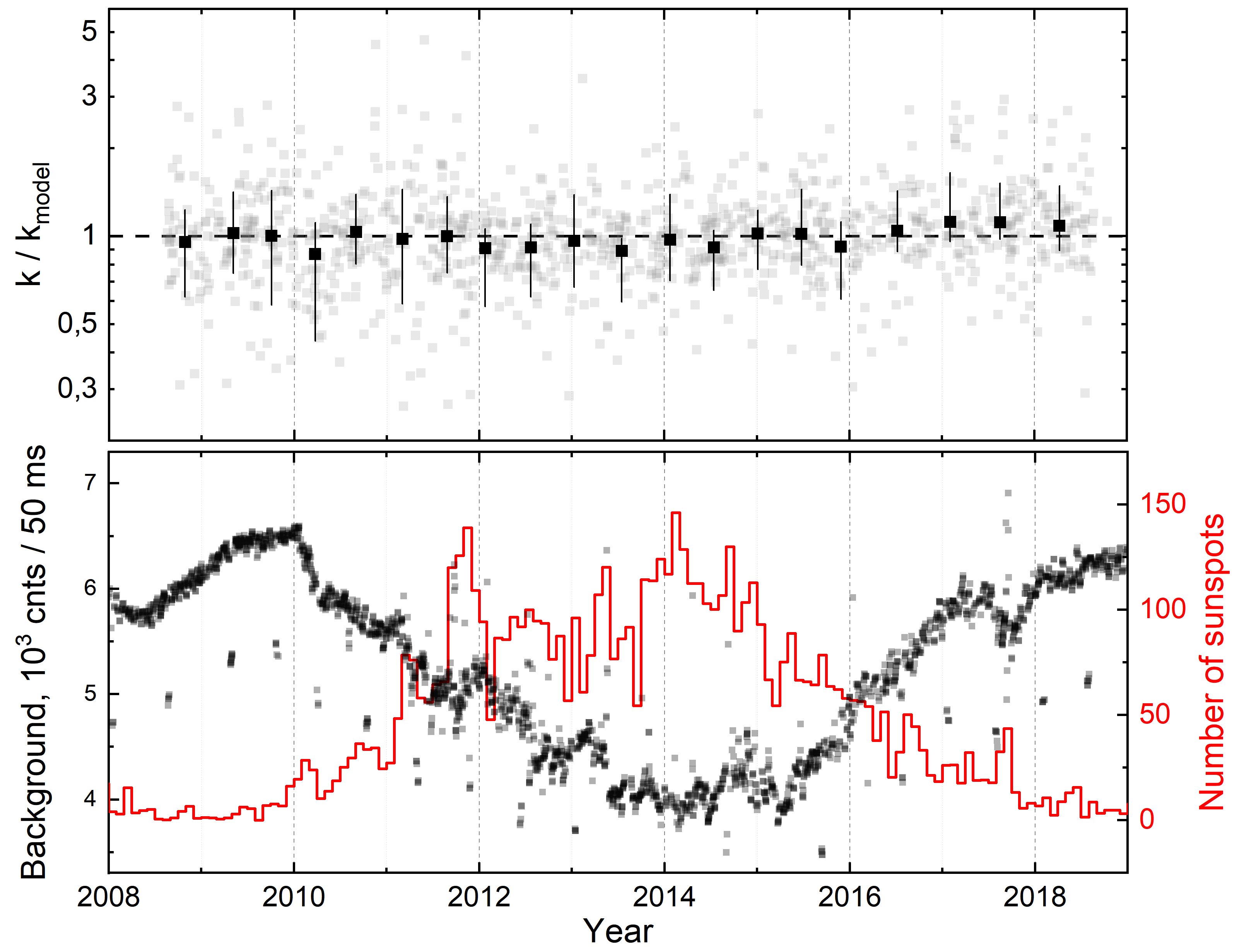}
    \caption{Dependence of corrected conversion factor $k/k_\text{model}$ (equation~\ref{eq:kmodel}) on the epoch of observations (top panel). Gray squares show individual events, black squares show median values for independent subsamples, each consisting of 50 events, registered successively. Error bars correspond to 68\% percentiles for the subsamples. Dashed horizontal line represents the absence of the evolution of the corrected conversion factor and its independence from the background level (see below). Corresponding dependence of SPI-ACS/INTEGRAL background level (black squares) and the number of sunspots (red curve) on the epoch are shown at bottom panel. }
    \label{fig:kevo}
\end{figure}

\subsubsection{Variance of the conversion factor}
\label{kvar}

Distribution of deviation of a measured conversion factor $k$ value from the analytical model $k_\text{model}$ for the whole sample of 1032 GRBs is shown in bottom left part of Fig.~\ref{fig:scatter}. The width (68\% percentile) of the distribution decreased by 1.7 times (in logarithmic scale) after all corrections, comparing with the initial distribution (top left part of Fig.~\ref{fig:scatter}). The distribution became  symmetric -- median and mean values of the distribution do not differ significantly (median($k/k_\text{model}$) = 0.98, mean($k/k_\text{model}$) = 0.96). 

The shape of the $k/k_\text{model}$ distribution is close to a Gaussian (Fig.~\ref{fig:scatter}), but with some excess at high deviations from the center of the distribution. It could be a superposition effect, explained by different width of the distribution for different sort of events. Indeed, the variance of conversion factor $k$ for GRBs with relatively soft energy spectrum (low $E_\text{p}$ value) should be higher because of low sensitivity of SPI-ACS at low energies (below 100 keV). 

We sort the whole sample of 1032 GRBs in ascending order of the $E_\text{p}$ parameter and form 10 independent subsamples, each containing of 100 events. For the subsamples we estimated widths of the corresponding $k/k_\text{model}$ distributions as 68\% percentiles,  defining values of standard deviation $\sigma_\text{k}$ of the conversion factor model. 

The dependence of the $\sigma_\text{k}$ values on the spectral hardness $E_\text{p}$ is shown in right part of Fig.~\ref{fig:scatter}. We find the correlation to be statistically significant: the Spearman rank-order correlation coefficient $\rho$ = -0.85 and the associated null-hypothesis (chance) probability $P_\rho$ = 1.6 $\times$ $10^{-3}$. The dependence is fitted by power-law model (equation~\ref{eq:sigm}, Table~\ref{ta_mopar}). 

\begin{equation}
  \sigma_\text{k} = E_1E_\text{p}^{~E_2}
	\label{eq:sigm}
\end{equation}

The parameter $\sigma_\text{k}$ defines an accuracy of our conversion procedure, it is estimated from the data directly and considers all possible uncertainties, connected with the calibration.

\subsubsection{Evolution of conversion factor with epoch of observations}

The sensitivity (effective area) of gamma-ray detectors could decrease with time due to various effects (mostly due to interaction of the equipment with galactic and solar cosmic rays). Our sample consists of more than one thousand events, registered continuously over ten years, which we can use to investigate possible evolution of the SPI-ACS sensitivity, measured as change in value of conversion factor $k$ (a higher value of $k$ corresponds to a lower sensitivity of the SPI-ACS).

We form 20 independent subsamples, each consisting of 50 events registered successively. For the subsamples we calculate median values of corrected conversion factor $k/k_\text{model}$ and corresponding 68\% percentiles, which we use as 1$\sigma$ error values. Results are presented in Fig.~\ref{fig:kevo}. 

We find no statistically significant evolution of conversion factor with the epoch of observations. Minor decrease of conversion factor in 2012-2015 and minor increase in 2016-2018 could be seen in Fig.~\ref{fig:kevo}. This behavior could be connected with the global evolution of background signal in SPI-ACS, anti correlating with the solar activity (measured as the number of sunspots), shown in the bottom part of Fig.~\ref{fig:kevo} (see also \citet{die18}).

Standard deviation of conversion factor median values for constructed subsamples (black squares in Fig.~\ref{fig:kevo}) defines an upper limit of conversion factor possible variation and evolution: the change in value of conversion factor is less than 7\%  (1$\sigma$ confidence level). In other words, we find possible change (loss) of SPI-ACS effective area with time to be no more than 7\%.

\subsection{SPI-ACS/INTEGRAL sensitivity for GRB-like transients}

\begin{figure}
	\includegraphics[width=\columnwidth]{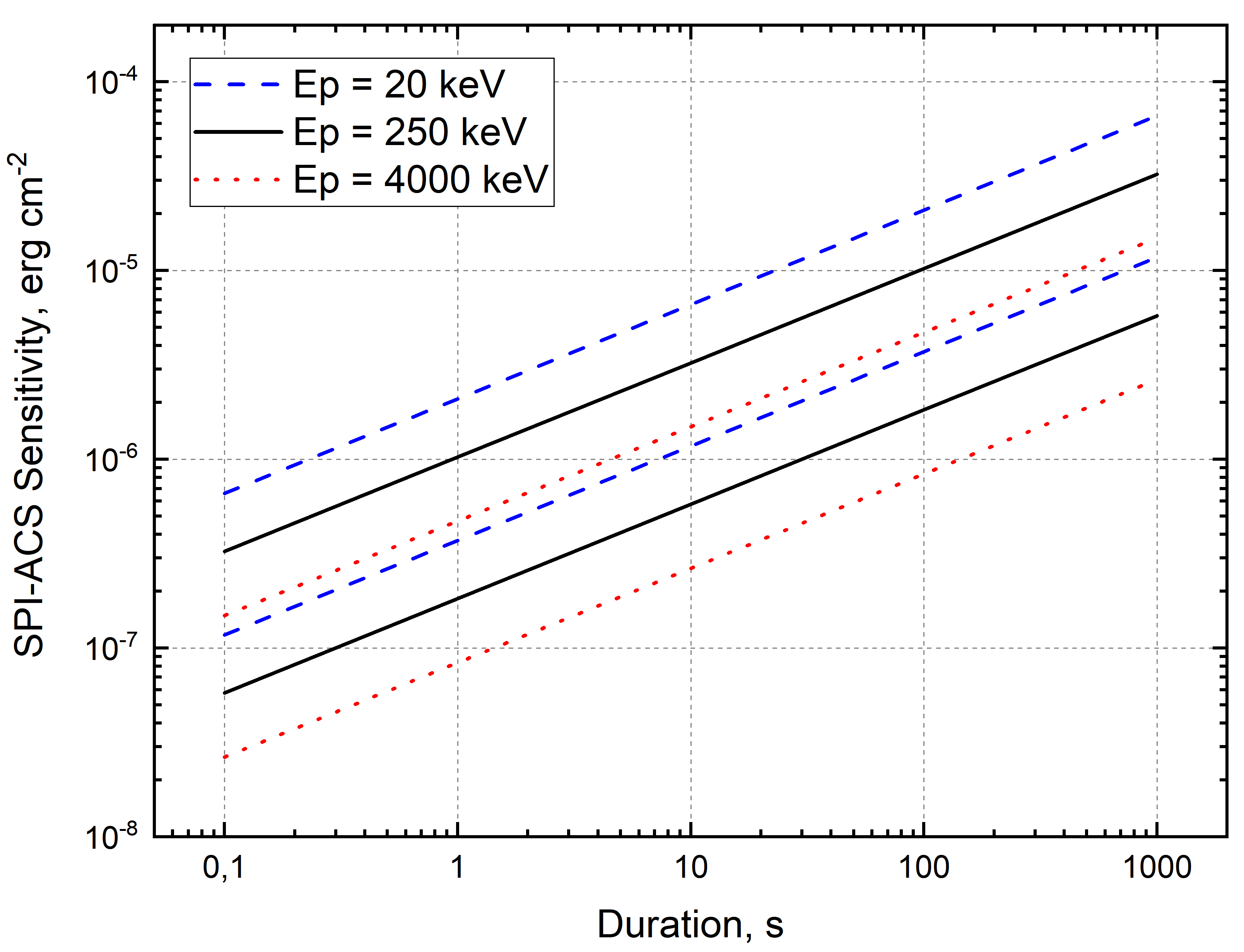}
    \caption{Dependence of SPI-ACS/INTEGRAL sensitivity (3$\sigma$ threshold over the background of $B$ = 5000 counts per 50 ms) on duration of the transient for three spectral models: $E_\text{p}$ = 20 keV (blue dashed lines), $E_\text{p}$ = 250 keV (black solid lines) and $E_\text{p}$ = 4000 keV (red dotted lines). The dependence for each model is shown by two curves, representing the most optimal and the least optimal detection angles.
    }
    \label{fig:sens}
\end{figure}

Using the model of conversion factor $k_\text{model}$, we estimate the sensitivity of SPI-ACS/INTEGRAL to detect transient sources with energy spectrum, analogous to gamma-ray bursts.

The sensitivity strongly depends on the level of background signal and its statistics. The long term evolution of SPI-ACS background is monotonic and anti-correlates with solar activity (Fig.~\ref{fig:kevo}). During investigated time period of 2008-2018, the background varies in wide range of (3900, 6500) counts per 50 ms (time resolution of SPI-ACS). In Jan 2023 typical value of SPI-ACS background is $B$ $\sim$ 5000 counts per 50 ms, which we use to estimate the sensitivity of SPI-ACS hereinafter. We provide all necessary information about the estimation, so one could make the calculations with any value of the background signal, if necessary. 

SPI-ACS is characterized by the deviation $d$ of the background $B$ from a Poissonian distribution: standard deviation 1$\sigma$ corresponds to $d$$\sqrt{B}$, where $d$ > 1 \citep{kien03, ryde03, rau05, min10a}. In \citet{min17} this effect is explained as a consequence of viewing of most BGO crystals by two PMTs (see Section~\ref{acs_descr}). 

We estimate the value of $d$ for the background in the light curve of each GRB from our sample and find the deviation $d$ to be dependent on the value of background signal. The dependence is linear, $d$ = $a$+$b$$\times$$B$, where $a$ = 1.31 $\pm$ 0.01, $b$ = (-3.2 $\pm$ 0.1) $\times$ 10$^{-5}$, and $B$ is background, measured in counts per 50 ms. Thus, for $B$ = 5000 counts per 50 ms we obtain $d$ = 1.15. 

We consider an event to be successfully detected when its fluence  exceeds threshold of 3$\sigma$: $S_\text{min}$ = 3$d$$\sqrt{B}$, where $B$ is the background signal, accumulated over the duration of the event. For example, for an event with duration of 1 s and $B$ = 5000 counts per 50 ms we obtain $S_\text{min}$ = 1090 counts, defining the sensitivity of SPI-ACS to transients at time scale of 1 s.   

Next, we estimate the sensitivity of SPI-ACS ($S_\text{min}$) in units of \ergc~in energy range of (10, 1000) keV, using our analytical model of conversion factor (equation~\ref{eq:kmodel}). For the event with duration of 1 s and $E_\text{p}$ = 500 keV, we obtain $S_\text{min}$ = 1.5 $\times$ 10$^{-7}$ \ergc~in case of optimal detection ($z$ $\simeq$ 70\degr~and $|a|$ $\ga$ 60\degr) and $S_\text{min}$ = 8.5 $\times$ 10$^{-7}$ \ergc~for $z$ $\simeq$ 120\degr~and $a$ $\simeq$ 0\degr.

Fig.~\ref{fig:sens} summarizes results of our SPI-ACS sensitivity estimation for background of $B$ = 5000 counts per 50 ms. It presents the dependence of sensitivity on the duration of the event for three models ($E_\text{p}$ = 20 keV, $E_\text{p}$ = 250 keV and $E_\text{p}$ = 4000 keV). The dependence for each model is shown by two curves, representing the most optimal and the least optimal detection. For example, at the shortest time scale of 0.1 s, $S_\text{min}$ = 2.6 $\times$ 10$^{-8}$ \ergc~for a GRB with $E_\text{p}$ = 4000 keV, detected in optimal conditions ($z$ $\simeq$ 70\degr~and $|a|$ $\ga$ 60\degr). At the same time, the sensitivity for a GRB with $E_\text{p}$ = 20 keV, registered in the least optimal direction ($z$ $\simeq$ 120\degr~and $a$ $\simeq$ 0\degr) is only $S_\text{min}$ = 6.6 $\times$ 10$^{-7}$ \ergc.

\subsection{Comparing our empirical calibration model with results of previous works}

\citet{vig09}, using the sample of 133 GRBs, found that 1 SPI-ACS count corresponds on average to $\sim$ 1 $\times$ $10^{-10}$\ergcc~in the (75, 1000) keV range, for directions orthogonal to the spacecraft pointing axis. We can not compare the calibration directly because of different energy range. Mean energy spectrum of the GRB sample in \citet{vig09} is described by Band model with $\alpha$ = -0.9, $\beta$ = -2.3 and $E_\text{p}$ = 450 keV, which corresponds according to our model to $k_\text{model}$ = 1.4 $\times$ $10^{-10}$\ergcc~at $z$ $\simeq$ 70\degr~and $|a|$ $\ga$ 60\degr~in energy range of (10, 1000) keV. Using the spectral model we estimate the conversion factor in (75, 1000) keV range, $k_\text{model}$ $\simeq$ 1.2 $\times$ $10^{-10}$\ergcc, which is marginally larger, than the value obtained in \citet{vig09}, but the difference is within the accuracy $\sigma_k$ of our calibration method.

Due to low number of events in the sample (133 GRBs), the detailed map of conversion factor and its dependence on spectral characteristics of bursts in \citet{vig09} were not developed. The dependence of conversion factor on the location of the source in spacecraft based coordinate system was investigated for three subsamples (top, central and bottom zones) divided according to the values of zenith angle. The conversion factor for top and bottom zones was found to be $\sim$ 2-3 times larger comparing with the central zone. Low sensitivity zone due to shielding of SPI-ACS by IBIS telescope was also found. 

Our analysis confirms these general tendencies in the dependence of conversion factor on the location of the source, but it shows the presence of additional zones, relatively small in angular size but very significant in impact on the value of conversion factor (Fig.~\ref{fig:model}).

We compare our detailed map of conversion factor with the map, estimated via Monte-Carlo simulations for LVT151012 event in \citet{sav17}, and find many similarities, including additional zones of low sensitivity at ($a$, $z$) = (0\degr, 25\degr) and ($a$, $z$) = (0\degr, 120\degr). The only difference is the 'ring' at $z$ $\simeq$ 150\degr~is not a 'ring' actually, but rather a half ring, placed at $|a|$ $\ga$ 90\degr, which we can not confirm due to low statistics of events in our sample in that area. 

The work \citet{poz20} was devoted to the study of short GRB 190425, registered by SPI-ACS/INTEGRAL. In \citet{poz20} the calibration was performed only for 278 relatively short ($T_{90}$ < 6 s) gamma-ray bursts, observed simultaneously by SPI-ACS/INTEGRAL and GBM/Fermi experiment. The restriction on the duration was chosen instead of estimation of spectral corrections to conversion factor. The dependence of conversion factor on the location of the source in spacecraft based coordinate system was not investigated. The calibration results obtained in \citet{poz20} are consistent with the results of this work, but give sufficiently larger uncertainties.

\section{Analysis of GRB/GW 190425}

\begin{figure*}
	\includegraphics[width=1.8\columnwidth]{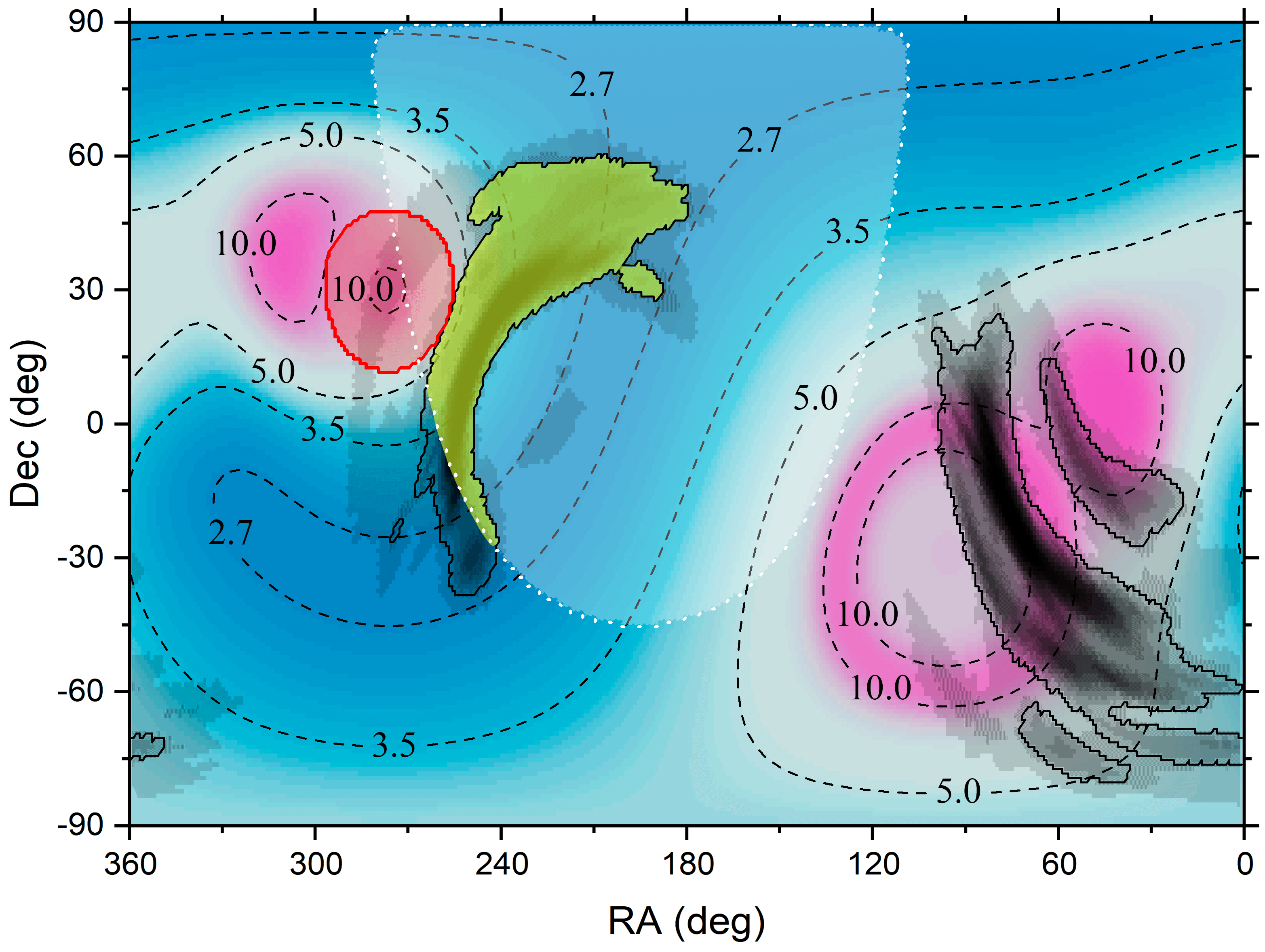}
    \caption{Localization map of GRB/GW 190425 event. Dependence of conversion factor $k_\text{model}$ on celestial coordinates for a GRB with characteristic energy of $E_\text{p}$ = 50 keV is shown by shades of blue-cyan-magenta with black dashed contour lines (labels of the contour lines show the values of $k_\text{model}$ in units of $10^{-10}$\ergcc). Localization map of gravitational wave event GW 190425 is shown by shades of gray with black solid contours at 68\% (1$\sigma$) significance level. The red circle (red solid contour + fill) shows the full field of view (35\degr) of INTEGRAL apertured telescopes. The area of the sky, covered by the Earth for the Fermi spacecraft at the moment of GW 190425 time trigger, is shown by white dotted line with light fill inside the area. The final localization area of GRB/GW 190425 event is shown by yellow fill. 
    }
    \label{fig:map}
\end{figure*}

In this section we apply our calibration model to the GRB/GW 190425 event, detected by SPI-ACS/INTEGRAL.

\subsection{Checking the model for GRB/GW 170817}
\label{sec:170817}

Before estimation of GRB 190425 energetic characteristics, we calculate the fluence of GRB 170817A (associated with gravitational wave event GW 170817) using SPI-ACS data and developed conversion factor model to compare the value with the fluence, obtained in GBM/Fermi experiment. It represents a final check of our model. 

According to \citet{poz20}, the fluence of the burst in SPI-ACS/INTEGRAL is $F_\text{SPI-ACS}$ = 570 $\pm$ 120 counts. The source of GRB 170817A is placed at ($a$, $z$) = (-68\degr, 105\degr) in spacecraft based coordinate system, close to the optimal area of detection. Spectral analysis results, obtained in \citet{poz18} using data of GBM/Fermi, give the fluence $F_\text{GBM}$ = (2.1 $\pm$ 0.3) $\times$ 10$^{-7}$\ergc~in (10, 1000) keV range and characteristic energy $E_\text{p}$ = 65$_{-14}^{+35}$ keV. 

Using equations~\ref{eq:kmodel},\ref{eq:sigm} we calculate conversion factor $k_\text{model}$ = (3.4$_{-1.1}^{+1.6}$) $\times$ 10$^{-10}$ \ergcc~for an event with ($a$, $z$) = (-68\degr, 105\degr) and $E_\text{p}$ = 65 keV. The converted fluence of GRB 170817A is $F_\text{SPI-ACS}$ = (1.9$_{-0.7}^{+1.0}$) $\times$ 10$^{-7}$ \ergc,  matching the $F_\text{GBM}$ value within 1$\sigma$ confidence region. 

The error of the converted fluence for GRB 170817A consists of two components, $\sigma_\text{F} = \sqrt{\sigma_\text{stat}^2 + \sigma_\text{sys}^2}$, where $\sigma_\text{stat}$ corresponds to statistical error of the fluence measurement in the SPI-ACS data and $\sigma_\text{sys}$ corresponds to the accuracy of calibration model (equation~\ref{eq:sigm}), which we call hereinafter as systematic error.

\subsection{The GRB/GW 190425 event}

Gravitational wave event GW 190425 is most likely the second binary neutron star merger in history, after GW 170817 \citep{abb17,abb20}. The possibility that one or both components of the binary system are black holes cannot be ruled out from gravitational-wave data. Total mass of this system is significantly larger than those of any other known binary neutron star system. It is sufficiently more distant event (comparing with GW 170817), placed at $D_\text{L}$ = 159$_{-71}^{+69}$ Mpc \citep{abb20}.

Nevertheless, weak GRB 190425, associated with gravitational wave event GW 190425, was discovered in \citet{poz20} in data of SPI-ACS/INTEGRAL. No confirmation by other gamma-ray experiments was obtained. The lack of detection of gamma-ray emission from GRB 190425 by the GBM/Fermi and BAT/Swift could be connected with  occultation of the source by the Earth. 

Estimates of GRB 190425 energetic characteristics in \citet{poz20} are based on simple calibration model, not taking into account localization of the source and its spectral characteristics. In the following sections we re-estimate energetic characteristics of GRB 190425, using our new analytical calibration model (equations~\ref{eq:kmodel},\ref{eq:sigm}).

\subsection{GRB/GW 190425 localization area}

The localization map of GW 190425 event, obtained in LIGO/Virgo experiments, is shown in Fig.~\ref{fig:map} by shades of grey with black solid contours at 68\% (1$\sigma$) significance level. The region is divided into two parts of nearly equal size.

The area, shielded by the Earth for Fermi observatory at the moment of GW 190425 trigger (white dotted contour with light fill in Fig.~\ref{fig:map}), covers most of north part of LIGO/Virgo localization map.

As GRB 190425 was not detected by both IBIS and SPI telescopes of INTEGRAL, the area, covering their field of view (red solid contour with fill in Fig.~\ref{fig:map}), could also be excluded from the localization area of GRB/GW 190425. The localization area of GW 190425 event at 68\% confidence level (black solid contours in Fig.~\ref{fig:map}) does not intersect with the IBIS and SPI field of views. 

Thus, the final localization area of GRB/GW 190425 event (yellow fill in Fig.\ref{fig:map}) covers the most of north part of LIGO/Virgo localization map, which is placed inside the Earth occultation area for Fermi and outside IBIS and SPI field of views.

\subsection{GRB 190425 fluence estimation}
\label{sec:190425flu}

For the moment of GW/GRB 190425 event, we construct corresponding map of SPI-ACS/INTEGRAL conversion factor $k_\text{model}$ (equation~\ref{eq:kmodel}) in equatorial coordinates, for different values of characteristic energy $E_\text{p}$, covering the range of (5, 10000) keV.

The conversion factor $k_\text{model}$ map for $E_\text{p}$ = 50 keV is shown in Fig.~\ref{fig:map} by shades of blue-cyan-magenta colors with black dashed contour lines. The localization area of GRB/GW 190425 (yellow fill) is placed in the zone of relatively high sensitivity of SPI-ACS/INTEGRAL experiment (shades of blue). All zones of low sensitivity (shades of magenta) do not intersect with the localization area of GRB/GW 190425. It simplifies estimates of GRB 190425 energetics, but does not cancel the necessity of introduction of the additional error component $\sigma_\text{map}$ of fluence, connected with variance of conversion factor model $k_\text{model}$ inside the localization area of GRB/GW 190425.

\begin{figure}
	\includegraphics[width=\columnwidth]{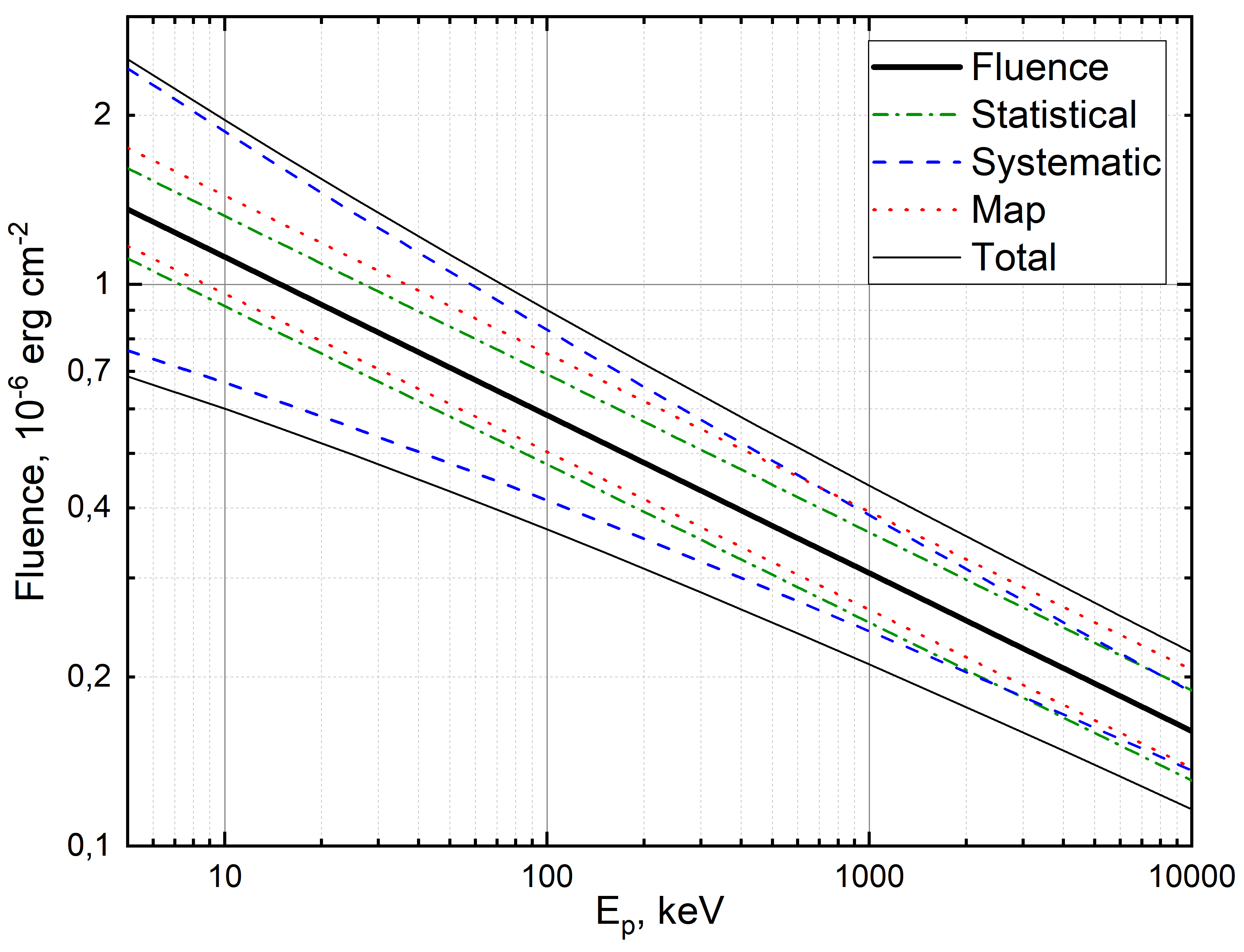}
    \caption{Dependence of GRB 190425 fluence in (10, 1000) keV energy range on characteristic energy $E_\text{p}$ (thick black line) with 1 $\sigma$ error intervals (thin black lines). Contribution of different error components is also shown: statistical by green dash-dotted lines, systematic (the width of $k/k_\text{model}$ distribution) by blue dashed lines and uncertainties from localization map (the width of $k_\text{model}$ distribution inside the localization area of GRB/GW 190425) by red dotted lines.
    }
    \label{fig:flu}
\end{figure}

The dependence of GRB 190425 fluence on characteristic energy $E_\text{p}$ is shown in Fig.~\ref{fig:flu}. We use median value of conversion factor model $k_\text{model}$ inside the localization area to estimate fluence. 68\% percentile of the $k_\text{model}$ distribution inside the localization area defines corresponding error $\sigma_\text{map}$ of fluence. Two other components of error (statistical $\sigma_\text{stat}$ and systematic $\sigma_\text{sys}$) are calculated analogously to the case of GRB 170817A (Section~\ref{sec:170817}). The total error is calculated via $\sigma_\text{F} = \sqrt{\sigma_\text{stat}^2 + \sigma_\text{sys}^2 + \sigma_\text{map}^2}$. As seen in Fig.~\ref{fig:flu}, $\sigma_\text{stat}$ $\simeq$ $\sigma_\text{map}$  $\simeq$ $\sigma_\text{sys}$ at high values of characteristic energy ($E_\text{p}$ $\ga$ 1000 keV), at low $E_\text{p}$ values $\sigma_\text{sys}$ component dominates (see equation~\ref{eq:sigm}). 

Possible range of GRB 190425 fluence covers almost one order of magnitude, varying from $F$ = 1.4 $\times$ $10^{-6}$ \ergc~for $E_\text{p}$ = 5 keV to $F$ = 1.6 $\times$ $10^{-7}$ \ergc~for $E_\text{p}$ = 10 MeV. In the following section we obtain possible constraints using the $ E_\text{p,i} $ -- $ E_\text{iso} $ correlation.

\subsection{GRB 190425 energetic constraints}

Gamma-ray bursts are characterized by a number of correlations between different observational parameters. We consider the correlation between the isotropic equivalent of the total energy emitted in the gamma-ray range, $E_\text{iso}$ and the characteristic energy in the source frame, $E_\text{p,i}$ \citep{ama02}. The nature of the correlation is still debatable, e.g. it could be connected with viewing angle effects \citep[e.g.][]{eic04,lev05,poz18}. It was shown, the $ E_\text{p,i} $ -- $ E_\text{iso} $ correlation could be used to classify GRBs and to estimate their redshift \citep[e.g.][]{min20a,min20b}. 

\begin{figure}
	\includegraphics[width=\columnwidth]{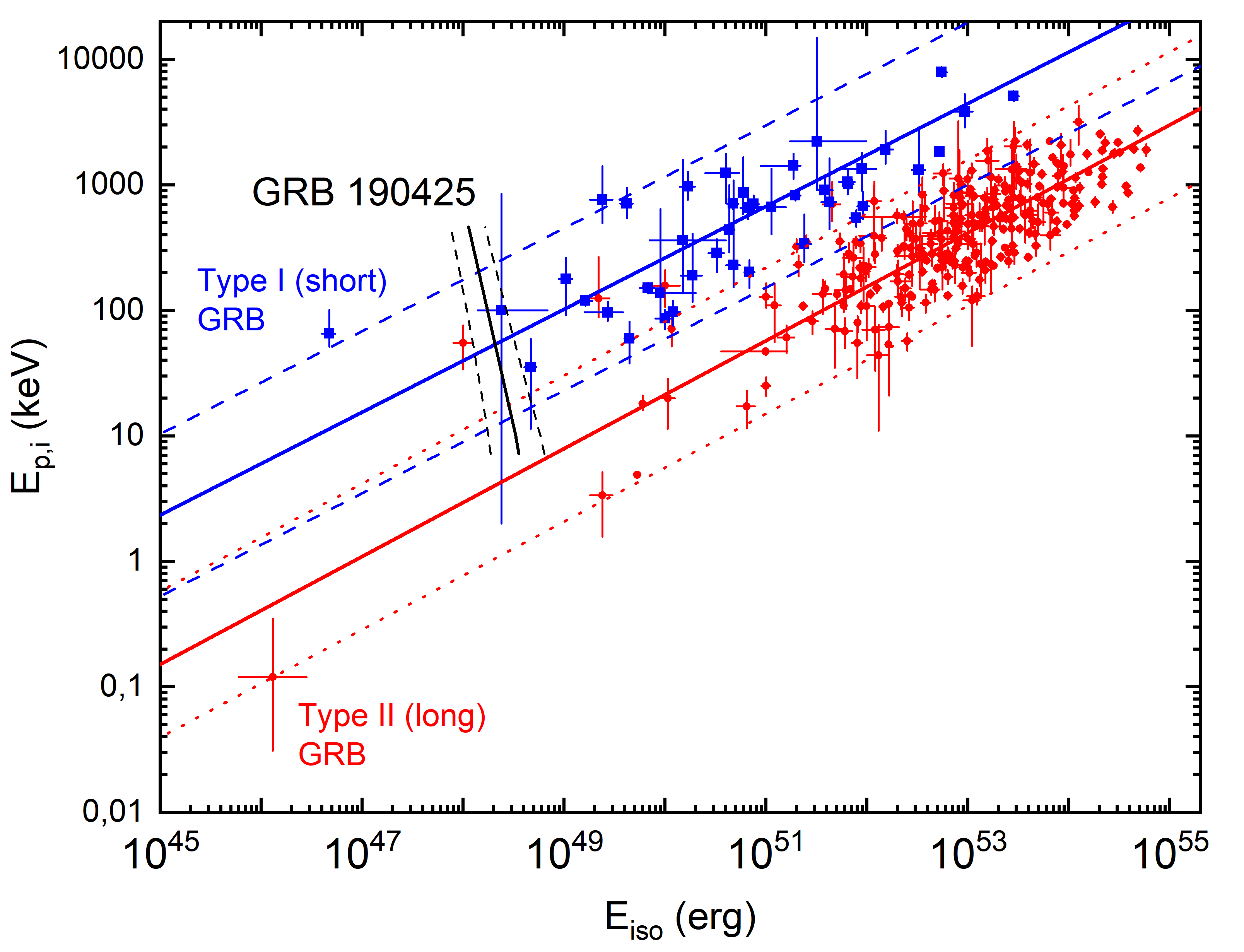}
    \caption{The $ E_\text{p,i} $ -- $ E_\text{iso} $ correlation of for type I (blue squares) and type II (red circles) GRBs with the approximation results (blue and red solid lines, respectively), including 2$\sigma_\text{cor}$ correlation regions (blue dashed and red dotted lines, respectively). The possible position of GRB 190425 (trajectory, depending on spectral properties of the event), assuming its classification as type I (short) burst, is shown by black solid line, black dashed lines cover 1$\sigma_\text{F}$ error interval.}
    \label{fig:amati}
\end{figure}

The correlation for the sample of 317 GRBs, taken from \citet{min20b,min21} is presented in Fig.~\ref{fig:amati}. As GRB 190425 is associated with neutron star merger, it should be classified as type I (short) burst. Using luminosity distance value of $D_\text{L}$ = 159 Mpc, obtained in LIGO/Virgo data \citep{abb20}, we calculate parameters $E_\text{p,i}$ and $E_\text{iso}$ for GRB 190425 fluence estimates from Section~\ref{sec:190425flu}. 

A trajectory of GRB 190425 on $ E_\text{p,i} $ -- $ E_\text{iso} $ diagram is shown in Fig.~\ref{fig:amati} by black solid line, black dashed lines correspond to 1$\sigma_\text{F}$ error interval. The trajectory crosses the correlation for type I bursts at $E_\text{p,i}$ = 55 keV and $E_\text{iso}$ = 2.0 $\times$ $10^{48}$ erg. We consider these values as the most probable for GRB 190425. 

To constrain possible range of  $ E_\text{p,i} $ and $ E_\text{iso} $ values, we use intersection points of the GRB 190425 trajectory with 2$\sigma$ correlation region. Thus, the constrained range of $E_\text{p,i}$ is (14, 200) keV and corresponding constrained range of $E_\text{iso}$ is (3.0, 1.4) $\times$ $10^{48}$ erg. This range is much narrower than previous estimates from \citet{poz20}. 

GRB 190425 is sufficiently more energetic than GRB 170817A: $ E_\text{iso} $ is larger by more than an order of magnitude, matching the faintest type I (short) GRBs of the sample from \citet{min20b,min21}. Larger $ E_\text{iso} $ could indicate a better orientation of the jet to the observer (tighter angle between jet axis and the line of sight) than for GRB 170817A, if the interpretation of the $ E_\text{p,i} $ -- $ E_\text{iso} $ correlation as viewing angle effect is correct \citep[e.g.][]{eic04,lev05,poz18}.

\section{Conclusions}

We perform cross-calibration of SPI-ACS/INTEGRAL experiment with GBM/Fermi, using data of 1032 bright gamma-ray bursts, detected by both experiments. 

Analytical model of conversion factor, $k_\text{model}$ is constructed (equation~\ref{eq:kmodel}). Firstly, it depends on spectral hardness of GRBs, defined by value of characteristic energy, $E_\text{p}$ (Fig.~\ref{fig:ecor}). Secondly, it depends strongly on a location of the source in spacecraft based coordinate system (Fig.~\ref{fig:ccor}). 

The dependence of the conversion factor on a location of the source is complex. Several zones of low sensitivity, connected with shielding of SPI-ACS by other instrumentation are found (Fig.~\ref{fig:model}).

We estimate the accuracy of our cross-calibration model empirically and found it to be dependent on spectral hardness parameter $E_\text{p}$ -- the accuracy is better for events with larger $E_\text{p}$ (Fig.~\ref{fig:scatter}). 

We find no statistically significant evolution of conversion factor with the epoch of observations in the period 2008 -- 2018, the possible change in value of conversion factor is less than 7\% at 1$\sigma$ confidence level (Fig.~\ref{fig:kevo}). 

Using the model of conversion factor, we estimate the sensitivity of SPI-ACS/INTEGRAL to detect gamma-ray bursts (Fig.~\ref{fig:sens}). For example, for an event with duration of 1 s and $E_\text{p}$ = 500 keV, minimal fluence (3$\sigma$ threshold) is $S_\text{min}$ = 1.5 $\times$ 10$^{-7}$ \ergc~in case of optimal detection angles.

We apply our calibration model to the GRB/GW 190425 event and re-estimate its energetic characteristics (Fig.~\ref{fig:flu}). Using the $ E_\text{p,i} $ -- $ E_\text{iso} $ correlation, we constrain possible range of them (Fig.~\ref{fig:amati}). We consider $E_\text{p,i}$ = 55 keV and $E_\text{iso}$ = 2.0 $\times$ $10^{48}$ erg values as the most probable for GRB 190425. 2$\sigma$ confidence level gives the range of (14, 200) keV for $E_\text{p,i}$ and the corresponding range of (3.0, 1.4) $\times$ $10^{48}$ erg for $E_\text{iso}$. This range is significantly narrower than previous estimates in \citet{poz20}.  

The calibration model could be applied to any transients with energy spectrum, analogous to one of gamma-ray bursts.

\section*{Acknowledgements}

Authors acknowledge support from Russian Science Foundation (RSF) grant 23-12-00198.

\section*{Data Availability}

The data underlying this article will be shared on reasonable request to the corresponding author.


\bibliographystyle{mnras}
\bibliography{PM}

\bsp	
\label{lastpage}
\end{document}